\documentclass[12pt,preprint]{aastex}
\slugcomment{}
\shorttitle{Lensed QSO SDSS~J1251+2935}
\shortauthors{}
\begin{document}
\title{A New Quadruply Lensed Quasar: SDSS~J125107.57+293540.5}
\author{
Issha Kayo\altaffilmark{1},
Naohisa Inada\altaffilmark{2,3},
Masamune Oguri\altaffilmark{4,5},
Patrick B. Hall\altaffilmark{6},
Christopher S. Kochanek\altaffilmark{7},
Gordon T. Richards\altaffilmark{8,9},
Donald P. Schneider\altaffilmark{10},
Donald G. York\altaffilmark{11},
and 
Kaike Pan\altaffilmark{12}
}

\altaffiltext{1}{Department of Physics and Astrophysics, Nagoya
  University, Chikusa-ku, Nagoya 464-8602, Japan.} 
\altaffiltext{2}{Institute of Astronomy, Faculty of Science, University
                 of Tokyo, 2-21-1 Osawa, Mitaka, Tokyo 181-0015, Japan.} 
\altaffiltext{3}{Cosmic Radiation Laboratory, RIKEN (The Physical and Chemical 
                 Research Organization), 2-1 Hirosawa, Wako, Saitama 351-0198, 
                 Japan.} 
\altaffiltext{4}{Kavli Institute for Particle Astrophysics and 
  Cosmology, Stanford University, 2575 Sand Hill Road, Menlo Park, 
  CA 94025, USA}
\altaffiltext{5}{Princeton University Observatory, Peyton Hall,
  Princeton, NJ 08544, USA}
\altaffiltext{6}{Department of Physics and Astronomy, York University,
4700 Keele Street, Toronto, Ontario, M3J 1P3, Canada}
\altaffiltext{7}{Department of Astronomy, The Ohio State University, 4055 McPherson Lab, 140 West 18th Avenue, Columbus, OH 43210, USA}
\altaffiltext{8}{Johns Hopkins University, 3400 N. Charles St., Baltimore, MD 21218, USA}
\altaffiltext{9}{Department of Physics, Drexel University, 3141 Chestnut Street, Philadelphia, PA 19104, USA}
\altaffiltext{10}{Department of Astronomy and Astrophysics, The
  Pennsylvania State University, University Park, PA 16802, USA}
\altaffiltext{11}{Department of Astronomy and Astrophysics, University
  of Chicago, Enrico Fermi Institute, 5640 South Ellis Avenue, Chicago,
  IL 60637, USA}
\altaffiltext{12}{Apache Point Observatory, New Mexico State Univ., P. O. Box 59, Sunspot, NM 88349, USA}

\begin{abstract}
We report the discovery of a quadruply imaged quasar,
SDSS~J125107.57+293540.5, selected from the Sloan Digital Sky Survey.
Follow-up imaging reveals that the system consists of four blue
point-like components in a typical cusp lens geometry surrounding a
central red galaxy. The source redshift is $0.802$ and the lens redshift
is $0.410$.  The maximum image separation between the lensed components
is $1\farcs79$.  While the image configuration is well reproduced by
standard mass models with reasonable parameter values, 
the flux ratios predicted by these models differ from
the observed ratios in all bands.  This is suggestive of small-scale
structures in this lens, although the definitive identification of the
anomaly requires more accurate photometry and astrometry.
\end{abstract}


\section{Introduction}\label{sec:intro}

Nearly one hundred gravitationally lensed quasars have been found to
date \citep[see the review by][]{kochanek06}, and they have become a
unique astronomical and cosmological tool.  Their usefulness is
particularly enhanced by constructing a statistical sample with a
well-understood selection function with the help of extensive and
homogeneous surveys such as the HST Snapshot Survey \citep{maoz93} or
the Cosmic Lens All Sky Survey \citep[CLASS;][]{myers03, browne03}.
However, the number of strongly lensed quasars in systematic surveys is
still limited.  To construct a larger lens sample, we are conducting the
SDSS Quasar Lens Search \citep[SQLS;][]{oguri+06} based on spectroscopic
quasar catalogs \citep{schneider05, schneider07} produced by the Sloan
Digital Sky Survey \citep[SDSS;][]{york00}.  Thus far, it has discovered
16 galaxy-scale and 2 cluster-scale lensed quasars
\citep{inada03a,inada03b,inada05,inada06a,inada06b,
inada07,johnston03,morgan03,pindor04,pindor06,oguri04,oguri05,morokuma07}
and re-discovered 8 previously known lensed quasars \citep{walsh79,
weymann80,surdej87,magain88,bade97,oscoz97,schechter98,morgan01}.  A
significant fraction of our lens candidates still require follow-up
observations, so we expect to discover more lensed quasars in the coming
years.

In this paper, we report the discovery of a quadruply lensed quasar,
SDSS~J125107.57+293540.5 (SDSS~J1251+2935), in the course of the SQLS.
Quadruple lenses, which constitute roughly one-third of all lensed
quasars, are not only visually interesting objects due to their
characteristic morphologies but also scientifically important and useful
objects.  Four image lenses provide many more constraints on the mass
distribution of the lens. In particular, the image flux ratios expected
for smooth, central potentials are relatively well-defined and the
differences between the observed and model flux ratios can be used to
study gravitational substructures \citep[e.g.,][]{mao98}.  Anomalous
flux ratios, where the observed and model image fluxes are in
significant disagreement are fairly common and are due to the combined
effects of the stars in the lens galaxy and satellite halos.  In radio
lenses this has been used to argue that the abundance of satellite halos
is consistent with the predictions of standard CDM models
\citep[e.g.,][]{metcalf01,chiba02,kochanek04}, while in
optical lenses the effects of substructure and microlensing by the stars
must be disentangled \citep[e.g.,][]{chiba05,keeton06,morgan06}.

In the next section, we briefly describe our algorithm for candidate
selection from the SDSS data.  Details of follow-up imaging using the
University of Hawaii 2.2-meter (UH88) telescope are provided in
\S~\ref{sec:image}. We investigate mass models of the system in
\S~\ref{sec:model}, and summarize our results in
\S~\ref{sec:sum}. Throughout the paper we assume a cosmological model
with matter density $\Omega_M=0.27$, cosmological constant
$\Omega_\Lambda=0.73$, and Hubble constant $h=H_0/(100 {\rm km}~{\rm
s}^{-1}{\rm Mpc}^{-1})=0.7$ \citep{spergel03}.

\section{SDSS Data}\label{sec:sdss}

The SDSS is conducting both a photometric survey
\citep{gunn98,lupton99,stoughton02,tucker06} in five broad-band optical
filters \citep{fukugita96} and a spectroscopic survey with a multi-fiber
spectrograph covering 3800{\,\AA} to 9200{\,\AA} at a resolution of
$\hbox{R}\sim1850$. The SDSS uses a dedicated wide-field ($3^{\circ}$
field of view) 2.5-m telescope \citep{gunn06} at the Apache Point
Observatory in New Mexico, USA, covering 10,000 square degrees of the
sky approximately centered on the North Galactic Cap. The imaging data
are processed by the photometric pipeline \citep{lupton01}, and
spectroscopic quasar targets are selected from the imaging data
according to the algorithm described by \cite{richards02}. Fibers for
the spectroscopic observations are assigned according to the tiling
algorithm of \citet{blanton03}. The imaging data have an astrometric
accuracy better than about $0\farcs1$ rms per coordinate \citep{pier03}
and photometric zeropoint errors less than about 0.03 magnitude over the
entire survey area \citep{hogg01,smith02,ivezic04}. SDSS~J1251+2935 is
contained in Data Release 5 \citep[for the imaging data;][]{adelman07}
and later (for the spectroscopic data).

SDSS~J1251+2935 was selected as a lensed quasar candidate using the
morphological selection algorithm described in \cite{oguri+06}.  The
algorithm uses the SDSS morphological classification parameter {\tt
objc\_type} and likelihood {\tt star\_L} that an object is fitted by a point-spread function (PSF).
Although lensed quasar systems with small image separations are
classified as single objects in the SDSS data, the profiles are extended
and are not consistent with either PSF profiles or simple galaxy
profiles.  Therefore, small separation lensed quasar candidates are
selected as objects that have very small values of {\tt star\_L}. In
addition, SDSS~J1251+2935 satisfies additional selection requirements
based on fits to the image with GALFIT \citep{peng02} which are applied
to exclude false positives by single quasars. While some single quasars
can pass the initial selection step, fits to such
systems using two PSFs lead to either very large magnitude differences
or very small image separations that are indicative of systematic errors
rather than a gravitational lens (see \S~5 of \cite{oguri+06}).

\begin{figure}
\epsscale{.7}
\plotone{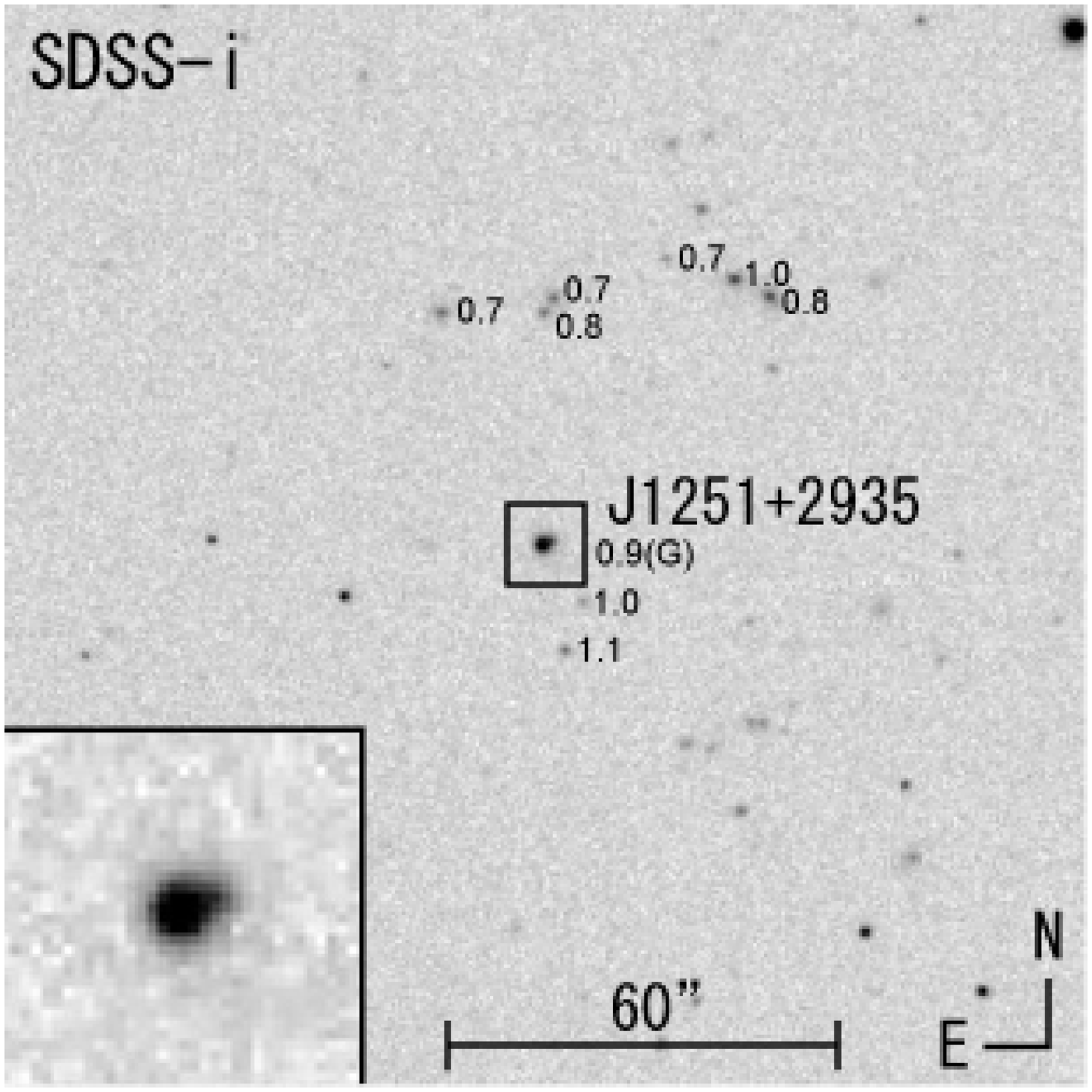}
\caption{
SDSS $i$-band image of the SDSS~J1251+2935 field (1\farcs2 seeing, 54 sec 
exposure). The pixel scale is $0\farcs396$ pixel$^{-1}$, North is up 
and East is left. An expanded view of SDSS~J1251+2935 is shown in the 
inset at lower left. Small numbers beside galaxies are $R-I$ colors from
 the Optic images.
\label{fig:sdssimg}}
\end{figure}

\begin{figure}
\plotone{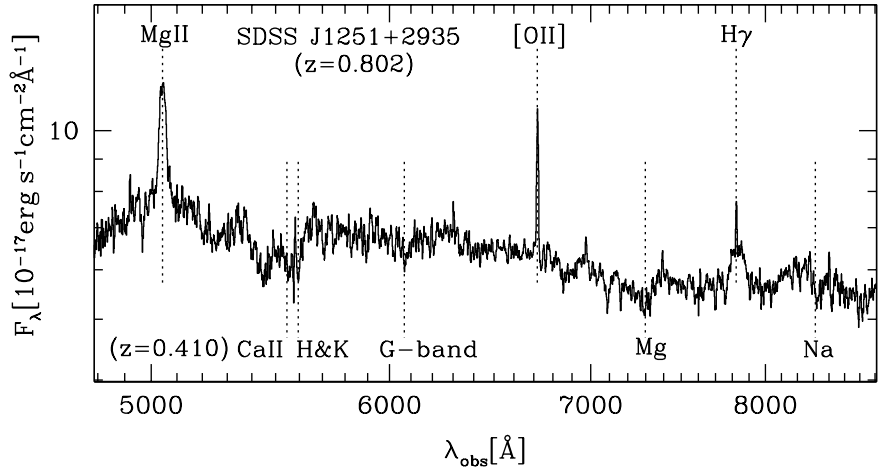}
\caption{
SDSS spectrum of SDSS~J1251+2935 with a resolution of $\sim1800$.
The quasar emission lines (\ion{Mg}{2}, [{\ion{O}{2}}], and H${\gamma}$)  
redshifted to $z=0.802$ are marked by the dashed lines and the galaxy 
absorption lines (Ca H\&K, G-band, Mg, and Na) redshifted to $z=0.410$ 
are marked by the dotted lines. The galaxy absorption lines indicate the
existence of a bright lensing galaxy.
\label{fig:spec}}
\end{figure}

The SDSS $i$-band image of SDSS~J1251+2935 is shown in
Figure~\ref{fig:sdssimg}.
The PSF magnitudes of SDSS~J1251+2935 (after
correcting for Galactic extinction) are $19.83\pm0.04$, $19.38\pm0.02$,
$19.13\pm0.03$, $18.85\pm0.03$, and $18.42\pm0.04$ in $u$, $g$, $r$,
$i$, and $z$, respectively.  SDSS~J1251+2935 is spectroscopically
confirmed as a quasar at $z=0.802$ (see Figure~\ref{fig:spec} for the
SDSS spectrum).  In this spectrum, we can also see a series of
absorption lines that indicate the presence of a bright early-type
galaxy at $z=0.410\pm0.001$.  The presence of both components is a
strong indication that this system is a gravitational lens.

\section{Imaging Follow-up Observations}\label{sec:image}

\begin{figure}
\epsscale{.85} \plotone{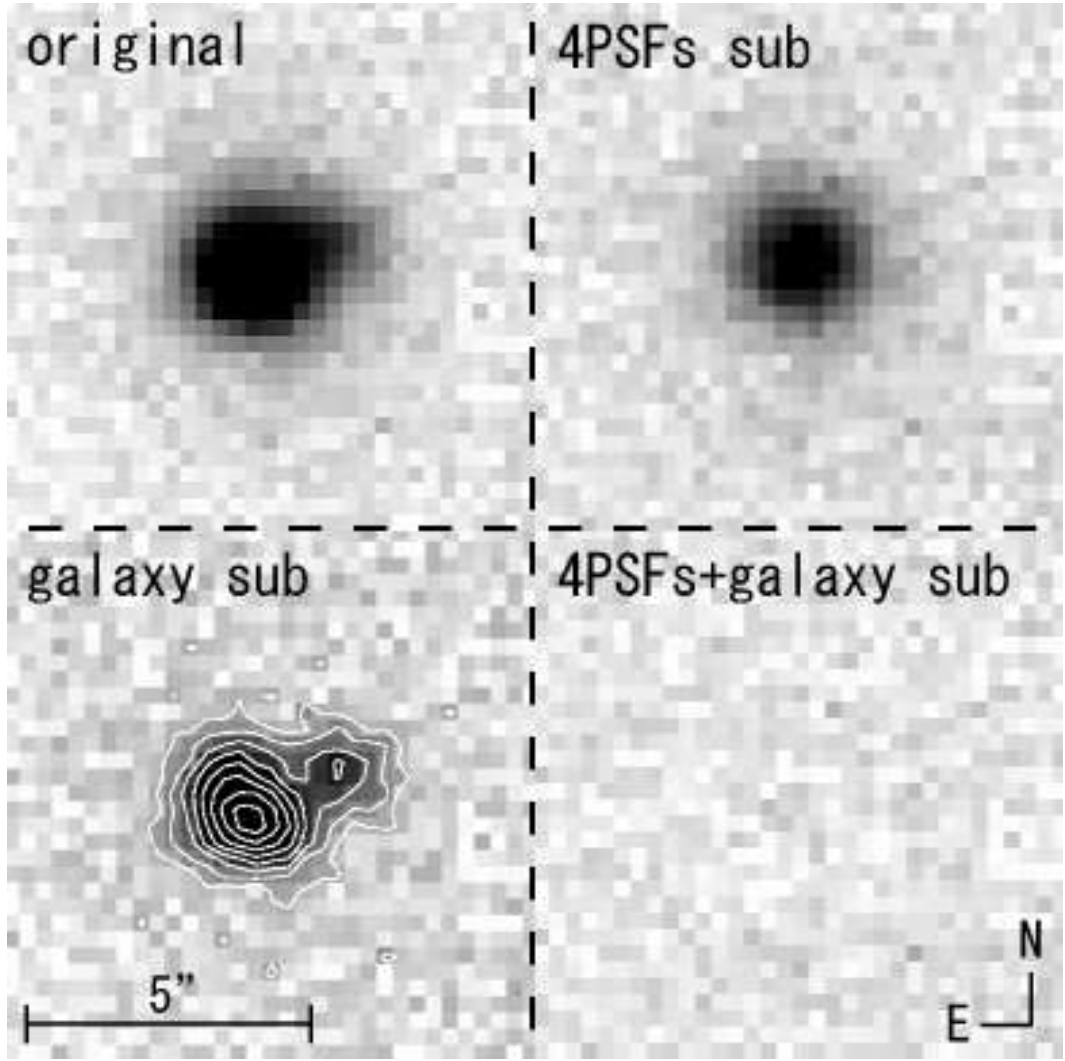}
\caption{Optic $I$-band image of SDSS~J1251+2935. The upper-left panel
is the original data. The galaxy-like extended object in the upper-right
panel is the residual after subtracting only 4 point-like components,
and vice versa in the lower-left (overplotted with contours). There are
no residuals after subtracting a galaxy and 4 PSFs, as shown in the
 lower-right panel. The images and contours are scaled by the
 square-root of the counts.\label{fig:image_sub}}
\end{figure}

\begin{figure}
\epsscale{.85}
\plotone{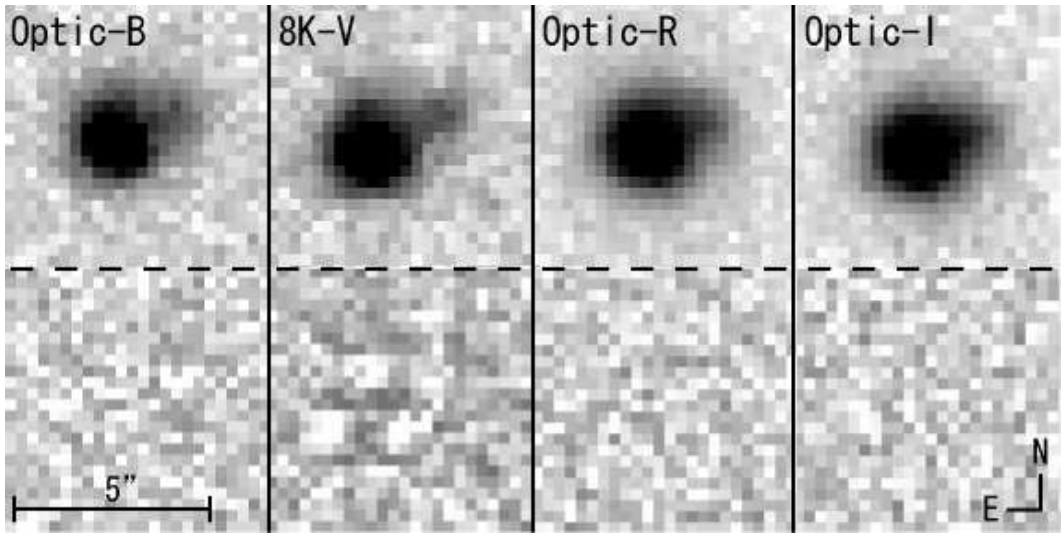} 
\caption{Original Optic
($BRI$) and UH8k ($V$) images of SDSS~J1251+2935 (upper panels) and the
residuals after subtracting the best models (lower panels).  The best
models consist of 4 PSFs for the Optic $B$-band and UH8k $V$-band,
and 4 PSFs plus a galaxy for Optic $R$ and $I$-bands.  The pixel scales of
Optic and UH8k are $0\farcs 274$ pixel$^{-1}$ (2${\times}$2 binned) and
$0\farcs 232$ pixel$^{-1}$, respectively. Although there are some 
residuals in the $V$-band image due to the galaxy, its flux is too small
 to be measured accurately.\label{fig:images}}
\end{figure}

Deeper and higher-resolution optical images of SDSS~J1251+2935 were 
obtained on 2006 April 25 (0\farcs8 seeing) and 2006 May 3 
(1\farcs0 seeing) using the 8k mosaic CCD camera (UH8k, pixel scale of 
$0\farcs232$ pixel$^{-1}$) and the Orthogonal Parallel Transfer Imaging 
Camera (Optic, pixel scale of $0\farcs137$ pixel$^{-1}$) at the UH88 
telescope, respectively. We took $V$- and $I$-band images (270 sec 
exposure for each band) with the UH8k, and $B$-, $R$-, and $I$-band 
images (400 sec exposure for each band) with the Optic. Because the 
night of UH8k imaging was not photometric, we did not photometrically
calibrate the UH8k images. We binned ($2\times2$) the Optic images
and used them for the astrometry and photometry of the system.
The $2\times 2$ binned Optic images ($BRI$) and the original UH8k image
($V$) are shown in the left panel of Figure~\ref{fig:image_sub} and the
upper panels of Figure~\ref{fig:images}.  

\begin{figure}
\plotone{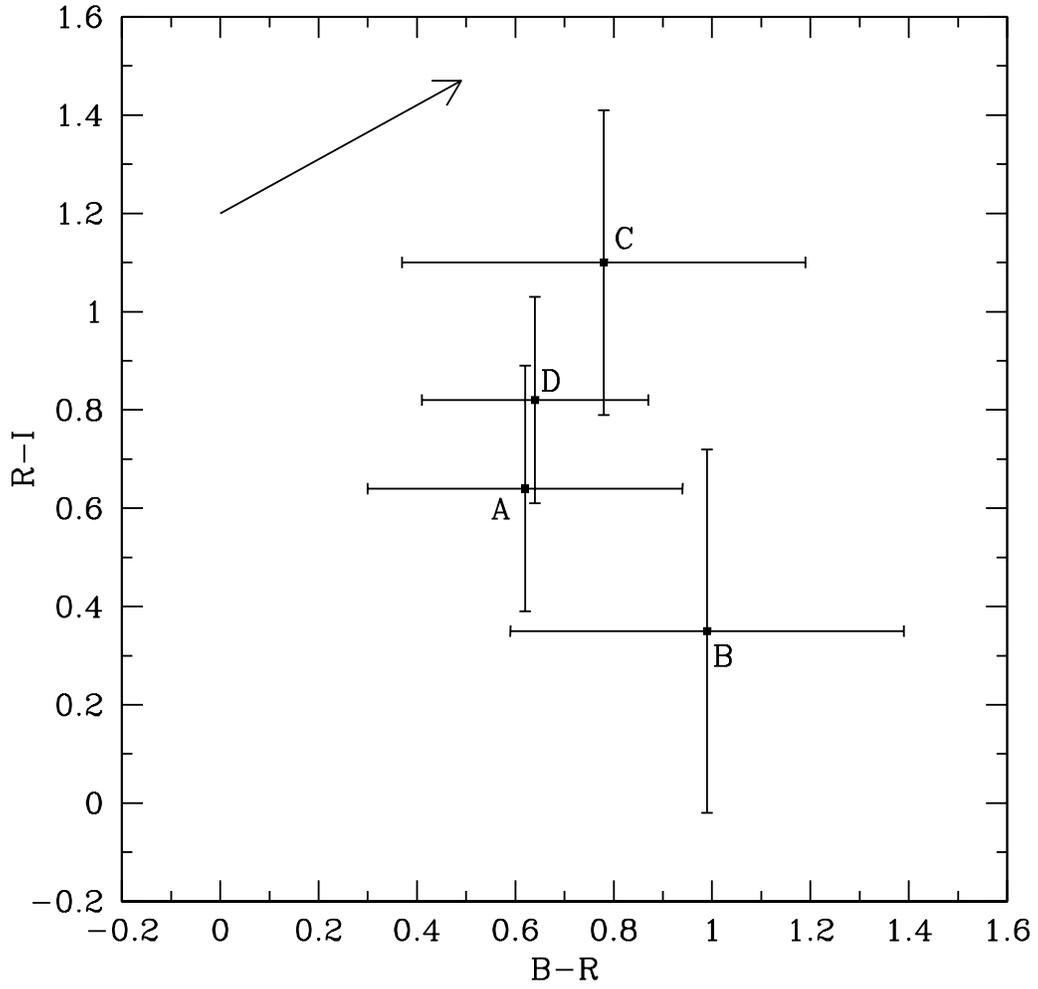}
\caption{
 Color-color diagram ($B-R$ and $R-I$)
 of the 4 point-like components.
The arrow upper-left indicates the extinction direction 
for a \citet{cardelli89} extinction law with $R_V=3.1$ and $\Delta(B-V)=0.3$.
\label{fig:color}}
\end{figure}

\begin{figure}
\epsscale{.5} \plotone{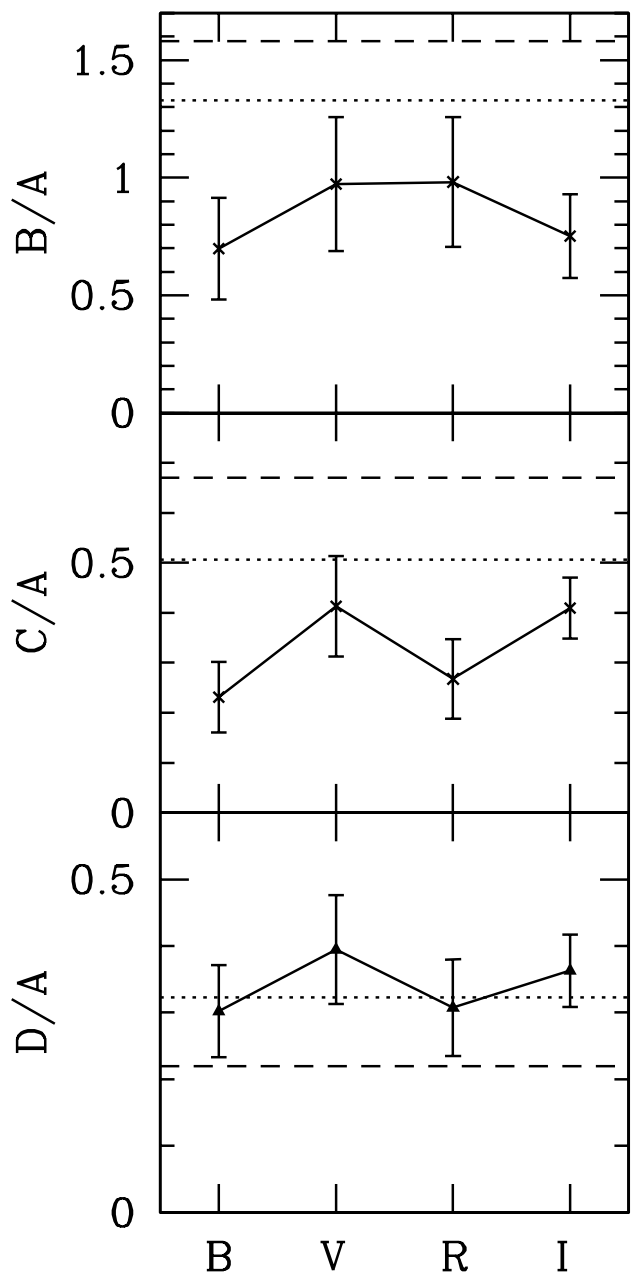}
\caption{ Flux ratios of the 4 point-like components.  The flux ratios 
predicted by the SIE model (without flux ratio constraints) and the
SIE plus external shear model (with the $R$-band flux ratio constraints
and the weak constraints on the ellipticity and the position angle of the
lens galaxy) in \S\ref{sec:model} are plotted as {\it dashed} and {\it
dotted} lines, respectively. Note that the latter model fits the
 observation with $\chi^2_{\rm red}=2.5$.  \label{fig:flux}}
\end{figure}

We used GALFIT to model these images with a series of models of
increasing complexity. The only model that works well consists of 4
point sources and a central galaxy modeled with a S\'{e}rsic profile.
In Figure~\ref{fig:image_sub}, we demonstrate that subtracting 4 fitted
PSFs from the Optic $I$-band image leaves an extended object or vice
versa.  If we further subtract the galaxy component, there remains
virtually no residual.  The lower panels of Figure~\ref{fig:images}
summarize the residuals after subtracting the best models for each
band's image.  The galaxy flux is well-determined only in the $R$ and
$I$-band images, so we neglected this component in the $B$ and $V$-band
fits even though there are hints of its presence in the $V$-band
residuals.  We label the 4 point sources A--D in order of increasing
$I$-band magnitudes and the central galaxy as G.  We estimated that the
galaxy has an effective radius of $1\farcs11\pm0\farcs38$, ellipticity
of $0.28\pm0.09$, the S\'{e}rsic index of $2.4\pm 1.1$, and a major axis
position angle of $26^\circ\pm 5^\circ$.  The $R-I$ color of G,
$0.89\pm0.30$, is consistent with an early-type galaxy at the measured
redshift \citep{fukugita95}.  Table~\ref{tbl:astro} summarizes the
relative astrometry and photometry of the system, where we defined the
errors from the scatter between the fits using 6 different PSF templates
rather than the smaller statistical uncertainties of the individual
fits.  The errors are an order of magnitude larger than other
systems where images are resolved even by visual inspection.  We plot the color-color
diagram ($B-R$ and $R-I$) of the 4 point-like components in
Figure~\ref{fig:color} and the flux ratios between the point-like
components in Figure~\ref{fig:flux}. Although the colors (and flux
ratios) of the lensed images have large scatter among the images, they
are consistent with each other given the large uncertainties.

In Figure~\ref{fig:sdssimg}, we also labeled the nearby galaxies by
their $R-I$ colors as measured using
SExtractor \citep{bertin96}.
There are several galaxies with colors similar to the lens galaxy to the
Southwest ($1.0<R-I<1.1$), and a larger group of generally bluer
galaxies ($0.6<R-I<1.0$) $\sim 40''$ to the North.
This suggests that the lens is associated with a group,
as is quite common among lensed quasars \citep[e.g.,][]{fassnacht02,
oguri05b, oguri06, williams06}.

\section{Mass Modeling}\label{sec:model}

\begin{figure}
\epsscale{1.0}
\plotone{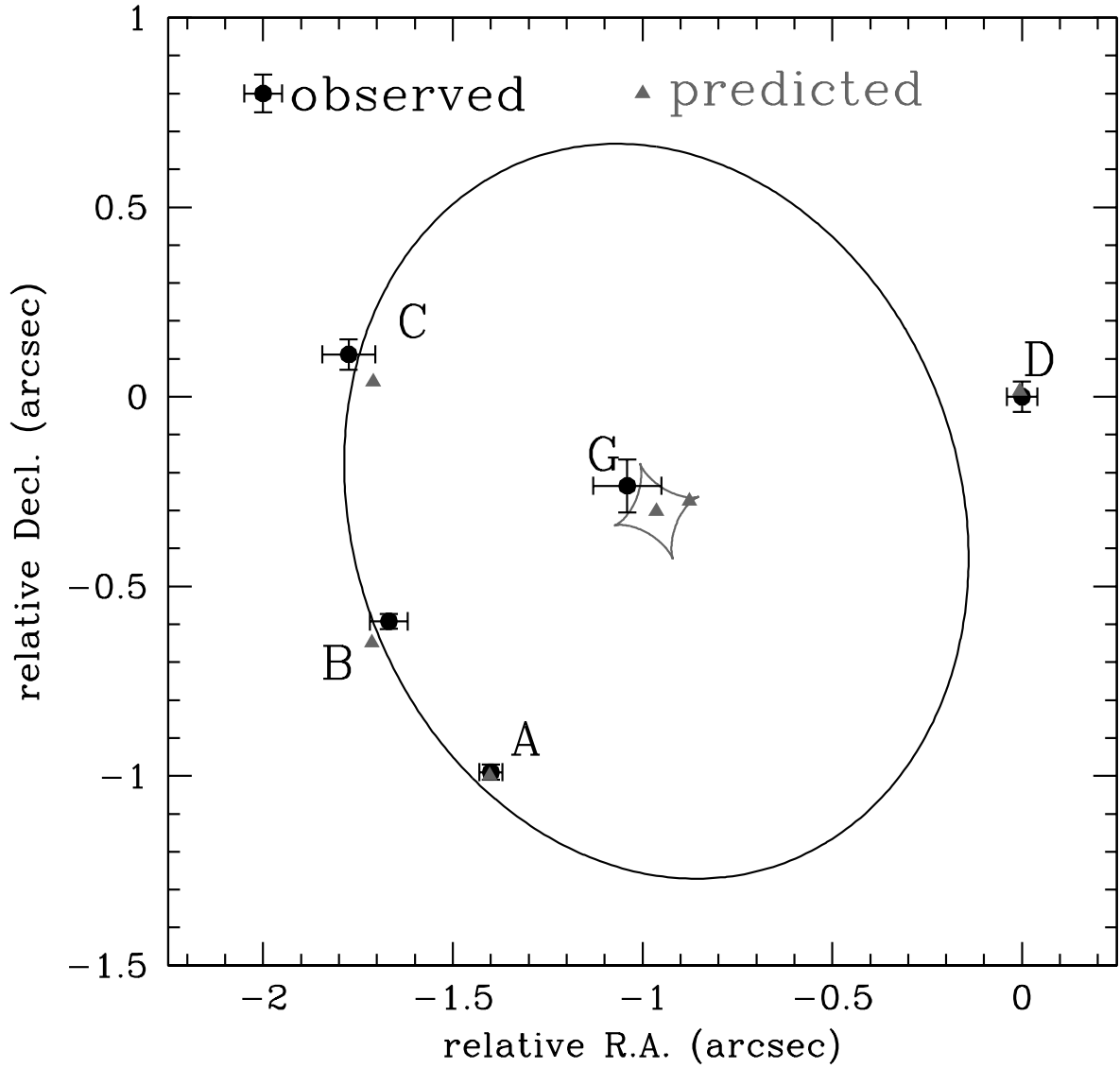}
\caption{ Predicted and observed positions of components A--D and G.
 The filled circles with error bars represents the observed positions,
 and the gray filled triangles represent the positions predicted by the
 SIE model.  The critical curve ({\it black line}) and the caustics
 ({\it gray line}) are also plotted. The gray filled triangle on the
 caustics is the predicted source plane position of the 
 quasar.  \label{fig:model}}
\end{figure}

We modeled the system using the {\it lensmodel} package \citep{keeton01}
to determine whether reasonable mass distributions can reproduce the
observations.  We first used the 7 parameter singular isothermal
ellipsoid model (SIE; the Einstein radius $R_{\rm E}$, the ellipticity
$e$ and its position angle $\theta_e$, and the positions of the galaxy
and the source quasar) to fit the relative positions of the A--D and G
components measured from the $R$-band image.  The model fits the data
well with $\chi^2_{\rm red}\equiv\chi^2/{\rm dof}=1.1$ for ${\rm dof}=3$
degrees of freedom.  The model parameters are presented in
Table~\ref{tbl:chi2} and the fit is illustrated in
Figure~\ref{fig:model}. The predicted position angle of the lens galaxy,
$19^\circ$, is reasonably consistent with the measured position angle of
$26^\circ\pm5^\circ$, which is typical for gravitational lenses
\citep{keeton98, koopmans06}. Based on the Einstein radius of the model,
the Faber-Jackson relations for gravitational lenses
measured by \citet{rusin03} predict a lens galaxy apparent magnitude of
$R=19.24$ ($I=18.57$) that agrees well with the measurement of
$R=19.32\pm 0.16$ ($I=18.43\pm0.25$).

Good fits to quadruple lenses generally require both the ellipticity of
the lens and an external shear \citep{keeton97}, so for our second model
we added a shear to the SIE model (the shear amplitude $\gamma$ and its
position angle $\theta_\gamma$).  In this case, the best fit model
overfits the data ($\chi^2_{\rm red}=0.062$ for ${\rm dof}=1$).
Although the resulting large amplitude of the external shear is not
inconsistent with N-body simulations \citep[e.g.,][]{holder03}, the lens
galaxy position angle is misaligned with respect to the observations.
Moreover, the large ellipticity and external shear cross almost
perpendicularly, which indicates that the model may not be
realistic.  If we add weak constraints to match the axis ratio and
position angle of the SIE component to the visible galaxy
($e=0.28\pm0.15$ and $\theta_e=26^\circ\pm10^\circ$), then we obtain a
good fit ($\chi^2_{\rm red}=1.2$ for dof=3) with a small external shear
($\gamma=0.02$).  This suggests that external shear is not important for
fitting the image positions despite the possible existence of nearby
groups.

In all these models, the predicted flux ratios differ from the observed
flux ratios (see Table~\ref{tbl:chi2} and Figure~\ref{fig:flux}).  This
remains the case if we add the flux ratios and their measurement errors
as model constraints.  The cusp relation \citep[see][]{keeton03}
provides a means of determining whether the flux ratios of the three
cusp images are consistent with any smooth mass distribution. In this
case we find $R_{\rm cusp}=0.13$ and $d/R_E=1.25$ based on the $R$-band
flux ratios and the SIE model, which is marginally consistent with the
range of distributions found for smooth lens models.  The origin of the
problem is presumably substructure in the gravitational potential of the
lens due to either microlensing by the stars or sub-halos,
since the observed flux ratios show no significant wavelength
dependence. We note, however, that evidence for anomalous flux ratios is
not conclusive, mainly because of large astrometric and photometric
errors.

\section{Summary}\label{sec:sum}

We report the discovery of the quadruply imaged quasar
SDSS~J1251+2935. The lensing hypothesis is confirmed by the facts that
i) the SDSS spectrum of the system shows the emission lines of a quasar
at $z_s=0.802$ and the absorption lines of a galaxy at $z_l=0.410$, ii)
the UH8k and Optic images confirm that the system consists of 4 blue
point-like components and an extended object whose color is consistent
with an elliptical galaxy at $z\sim 0.4$, iii) the geometry of the
system is that of a typical cusp lens and can be well reproduced by a
SIE model with reasonable parameter values, and iv) the luminosity of
the lens galaxy is consistent with the expected luminosity from the
Faber-Jackson relation.  This system is the second lowest redshift
lensed quasar after RXJ1131-1231 \citep{sluse03} at $z_s=0.66$. 
The flux ratios of the three cusp images of the lens show a modest flux
ratio anomaly, whose origins could be better constrained with higher
resolution images under better observing conditions.

\acknowledgments

We thank Atsunori Yonehara for many useful comments and the anonymous
referee for suggestions to improve the manuscript.
Use of the UH 2.2-m telescope for the observations is supported by NAOJ.
I.~K. acknowledges the support from Ministry of Education, Culture,
Sports, Science, and Technology, Grant-in-Aid for Encouragement of Young
Scientists (No. 17740139).  
N.~I. acknowledges supports from the Japan Society for the Promotion of 
Science and the Special Postdoctral Researcher Program of RIKEN. 
This work was supported in part 
by Department of Energy contract DE-AC02-76SF00515.

Funding for the creation and distribution of the SDSS Archive has been
provided by the Alfred P. Sloan Foundation, the Participating
Institutions, the National Aeronautics and Space Administration, the
National Science Foundation, the U.S. Department of Energy, the Japanese
Monbukagakusho, and the Max Planck Society. The SDSS Web site is
http://www.sdss.org/.

The SDSS is managed by the Astrophysical Research Consortium for the
Participating Institutions. The Participating Institutions are the
American Museum of Natural History, Astrophysical Institute Potsdam,
University of Basel, Cambridge University, Case Western Reserve
University, University of Chicago, Drexel University, Fermilab, the
Institute for Advanced Study, the Japan Participation Group, Johns
Hopkins University, the Joint Institute for Nuclear Astrophysics, the
Kavli Institute for Particle Astrophysics and Cosmology, the Korean
Scientist Group, the Chinese Academy of Sciences (LAMOST), Los Alamos
National Laboratory, the Max-Planck-Institute for Astronomy (MPIA), the
Max-Planck-Institute for Astrophysics (MPA), New Mexico State
University, Ohio State University, University of Pittsburgh, University
of Portsmouth, Princeton University, the United States Naval
Observatory, and the University of Washington.

\clearpage

\begin{deluxetable}{crrrrrr}
\rotate
\tablewidth{0pt}
\tablecaption{ASTROMETRY AND PHOTOMETRY OF SDSS~J1251+2935\label{tbl:astro}}
\tablehead{\colhead{Object} & \colhead{$x$[arcsec]\tablenotemark{a}} &
 \colhead{$y$[arcsec]\tablenotemark{a}} &  
 \colhead{$B$\tablenotemark{b}} &
 \colhead{$R$\tablenotemark{b}} &
 \colhead{$I$\tablenotemark{b}}} 
\startdata
A & $-1.40\pm0.03$ & $-1.00\pm0.02$
 & $20.67\pm0.22$  & $20.05\pm0.23$ & $19.41\pm0.10$ \\
B & $-1.67\pm0.05$ & $-0.65\pm0.02$
 & $21.06\pm0.31$  & $20.07\pm0.25$ & $19.72\pm0.27$ \\
C & $-1.77\pm0.07$ & $0.04\pm0.04$
 & $22.26\pm0.30$  & $21.48\pm0.28$ & $20.38\pm0.14$ \\
D & $0.00\pm0.04$  & $0.00\pm0.04$
 & $21.97\pm0.16$  & $21.33\pm0.16$ & $20.51\pm0.14$  \\
G & $-1.04\pm0.09$ & $-0.23\pm0.07$
 & \nodata & $19.32\pm0.16$ & $18.43\pm0.25$ \\
\enddata
\tablenotetext{a}{Measured in the Optic $R$-band image using GALFIT. The
 positive directions of $x$ and $y$ are West and North,
 respectively. Errors indicate the dispersions from 6 different PSF
 templates.}  \tablenotetext{b}{Measured in the Optic images using
 GALFIT. The errors are the dispersions from 6 different PSF templates,
 and they do not include the absolute calibration uncertainties. The
 magnitudes are calibrated using the standard star PG0918+029
 \citep{landolt92}.}
\end{deluxetable}

\begin{deluxetable}{llrrcccccl}
\rotate
\tablewidth{0pt}
\tablecaption{SDSS~J1251+2935: MASS MODELS\label{tbl:chi2}}
\tablehead{\colhead{Model} &
 \colhead{Data\tablenotemark{a}} &
 \colhead{$\chi^2_{\rm tot}$/dof\tablenotemark{b}} &
 \colhead{$\chi^2_{\rm flux}$\tablenotemark{c}} &
 \colhead{$R_{\rm E}$[arcsec]} &
 \colhead{$e$} &
 \colhead{$\theta_e$[deg]} &
 \colhead{$\gamma$} &
 \colhead{$\theta_\gamma$[deg]} &
 \colhead{comments}}
\startdata
SIE        & pos           & 3.3/3 & (20) & 0.88 & 0.19 & 19
 & \nodata & \nodata
 & bad flux\\
SIE+shear  & pos           & 0.062/1 & (18) & 0.79 & 0.66 & 10
 & 0.21 & $-80$ 
 & large misalignment \\
SIE+shear  & pos+shape      & 3.7/3 & (16) & 0.87 & 0.21 & 25 
 & 0.018 & $-8.7$ 
 & bad flux \\
SIE        & pos+flux      & 17/6 & 8.8   & 0.80 & 0.46 & 18
 & \nodata & \nodata
 & poor fitting \\ 
SIE+shear  & pos+flux      & 8.9/4 & 1.1  & 0.76 & 0.67 & 53
 & 0.28 & $-32$ 
 & large misalignment \\
SIE+shear  & pos+flux+shape & 15/6 & 6.3   & 0.83 & 0.38 & 34
 & 0.074 & $-15$ 
 & poor fitting \\ 
\enddata
\tablecomments{Results of various mass models constrained by $R$-band
 data. The position angles are measured East of North. The time
 delay between images A and D in the SIE model is $\Delta t_{AD}\sim 17$
 $h^{-1}$day.}
\tablenotetext{a}{Data used to constrain the models; pos: positions of
 the 4 images and the galaxy, flux: fluxes of the 4 images, and
 shape: weak constraints on the ellipticity and position angle of 
 the lens galaxy ($e=0.28\pm0.15$ and $\theta=26^\circ\pm10^\circ$).}
 \tablenotetext{b}{Total $\chi^2$ and the degree of freedom.}
 \tablenotetext{c}{Contribution of fluxes to the $\chi^2$. The values
 in parentheses are not included in $\chi^2_{\rm tot}$.}
\end{deluxetable}

\end{document}